\documentclass[showpacs,prd]{revtex4}

\usepackage{epsfig}
\usepackage{graphicx}
\usepackage{dcolumn}
\usepackage{amsmath}
\usepackage{latexsym}

\begin{document}

\title{Classical dynamics on three dimensional fuzzy space: Connecting the short and long length scales}  
\author{FG Scholtz$^{a,b}$ }
\affiliation{$^a$National Institute for Theoretical Physics (NITheP), 
Stellenbosch 7602, South Africa\\
$^b$Institute of Theoretical Physics, 
Stellenbosch University, Stellenbosch 7602, South Africa\\}

\begin{abstract}
\noindent 
 We derive the path integral action for a particle moving in three dimensional fuzzy space.  From this we extract the classical equations of motion.  These equations have rather surprising and unconventional features: They predict a cut-off in energy, a generally spatial dependent limiting speed, orbital precession remarkably similar to the general relativistic result, flat velocity curves below a length scale determined by the limiting velocity and included mass, displaced planar motion and the existence of two dynamical branches of which only one reduces to Newtonian dynamics in the commutative limit.  These features place strong constraints on the non-commutative parameter and coordinate algebra to avoid conflict with observation and may provide a stringent observational test for this scenario of non-commutativity.

\end{abstract}

\maketitle



\section{Introduction}
\label{intro}

The structure of space-time at short length scales and the emergence of space-time as we perceive it at long length scales are probably the most challenging problems facing modern physics \cite{seib}.  These issues are also at the core of the struggle to combine gravity and quantum mechanics into a unified theory and probably also links closely with the observational challenges of dark matter and energy.

One of the difficulties facing our understanding of space-time at short length scales is the lack of observational data that can be accessed at energies and length scales available to us, either through accelerators or astronomical observation.  Mostly the short length scale structure of space time manifests itself at very high energies and short length scales inaccessible to current observational techniques.  

This lack of understanding how the short length scale structure of space-time may imprint itself at long length scales and low energies is what motivates the current paper.  What we aim to achieve in this paper is to show that one scenario for space time at short length scales, namely, non-commutative space and more specifically fuzzy space, has very clear observational consequences at the solar and galactic scales that severely constrain the non-commutative coordinate algebra.  

Non-commutative space-time has received considerable attention in the past few decades.  This was originally proposed by Snyder \cite{Snyder} in an attempt to avoid the ultra-violet infinities of field theories.  The discovery of renormalization pushed these ideas to the background until more recently when they resurfaced in the search for a consistent theory of quantum gravity.  The compelling arguments of Doplicher et al \cite{dop} highlighted the need for a revised notion of space-time at short length scales and gave strong arguments in favour of a non-commutative geometry.  Shortly thereafter it was also noted that non-commutative coordinates occurred quite naturally in certain string theories \cite{wit}, generally perceived to be the best candidate for a theory of quantum gravity.  This sparked renewed interest in non-commutative space-time and the formulation of quantum mechanics \cite{scholtz} and quantum field theories on such spaces \cite{doug}.  

Despite the developments above, the observational consequences of non-commutativity remain elusive due to the smallness of the effect, especially on the microscopic level.  In \cite{scholtz1,scholtz2, scholtz3} it was argued that non-commutativity can have observational consequences at the macroscopic scale for Fermi gases at very high densities and/or temperatures.  Yet, again, if non-commutativity is assumed to manifest itself at the Planck scale, these densities and temperatures are outside our observational window.  

One other possible manifestation of non-commutativity on the macroscopic level may be in the modification of classical dynamics and gravity.  This has for many years been the topic of what is generally referred to as modified Newton dynamics (MOND) \cite{clif} and modified gravity \cite{noj}.  Although the present paper contains elements of MOND, there are many technical differences.  Furthermore, in contrast to the phenomenological approach of MOND, the modified dynamics derived here follow from first principles using the action of a particle moving in a non-commutative space.  The latter, in turn, is also systematically derived from the Schr\"{o}dinger equation.   

As our aim here is to study the motion of macroscopic objects, including planetary and galactic motion, it is sufficient to limit ourselves to the non-relativistic regime of low velocities.  The starting point of our derivation will therefore be the non-relativistic Schr\"{o}dinger equation.  There are, of course, effects arising in planet motion that stem from a general relativistic description.  The prime example is precession of the perihelion of a planet's orbit such as Mercury.  We show here that non-commutative dynamics gives rise to the same effect, albeit numerically somewhat different.  Not only this, but another feature of relativity that seems to emerge quite naturally is that of a limiting speed.  

This paper is organised as follows: In section \ref{section 1} we give a generic expression for the action of a particle on two and three dimensional non-commutative space.  In section \ref{section 2} we specify to a two dimensional non-commutative plane and derive the modified equations of motion. In section \ref{section 3} we consider three dimensional  fuzzy space and derive the modified equations of motion.  Due to its observational importance, we spend considerable time discussing the implications of this modified dynamics in a number of subsections.  In section \ref{geneqm}, we introduce a further generalization of the modified dynamics found in section \ref{section 3}.  In section \ref{discussion} we discuss the core findings, their implications and open issues.  Finally, we close with a summary and conclusions in section \ref{summary}

\section{The action on non-commutative space}
\label{section 1}

In non-commutative space coordinates are no longer commuting and simultaneous eigenstates of position cannot be found.  This eliminates the standard time slicing procedure using position eigenstates and the subsequent representation of the transition amplitude as a functional integral over position. Instead, one needs to replace the eigenstates of position with minimum uncertainty states, which also form an overcomplete set, commonly referred to as coherent states.  The standard time slicing procedure can then again be implemented to write a coherent state path integral representation of the transition amplitude.  In general this representation reads  \cite{klauder}
\begin{equation}
\langle\ell_f,t_f|\ell_i,t_i\rangle=\int_{\ell(t_i)=\ell_i}^{\ell(t_f)=\ell_f}[d\mu(\ell)] e^{\frac{i}{\hbar} S},
\end{equation}
with the path integral action
\begin{equation}
\label{action}
S=\int_{t_i}^{t_f}dt \langle\ell(t)|i\hbar\frac{\partial}{\partial t}-H|\ell(t)\rangle.
\end{equation}
Here $|\ell\rangle$ is a set of overcomplete coherent states, i.e.
\begin{equation}
\label{overc}
\int d\mu(\ell) |\ell\rangle\langle\ell |=\bf{1},
\end{equation} 
with $\bf{1}$ the identity on the Hilbert space.  

This is the strategy we employ here to derive the path integral action for a particle on non-commutative space.  In the next section we derive the action and classical dynamics of a particle in the two dimensional non-commutative plane.

\section{The two dimensional non-commutative plane}
\label{section 2}

To start, we briefly recall the formulation of quantum mechanics on two dimensional non-commutative space \cite{scholtz}.  In this case the coordinate algebra is given by
 \begin{equation}
\label{ncom2d}
\left[\hat{x} , \hat{y} \right] = i \theta
\end{equation}
where $\theta$ is a constant with dimension of a length squared, which we can take without loss of generality to be positive,  and $\hat{x}$, $\hat{y}$ are hermitian operators.

To develop the quantum theory \cite{scholtz}, one first introduces a representation of this coordinate algebra on some Hilbert space ${\cal H}_c$, referred to as classical configuration space.  In the case at hand, one notes that $b=\frac{1}{\sqrt{2\theta}}(\hat{x}+i\hat{y})$ and $b^\dagger=\frac{1}{\sqrt{2\theta}}(\hat{x}-i\hat{y})$ are standard creation and annihilation operators.  The radius operator is $\hat{r}^2=\hat{x}^2+\hat{y}^2=\theta (b^\dagger b+1)$.  It is then natural to choose for ${\cal H}_c$ the Fock space for one oscillator \cite{scholtz} since each value of the quantised radius appears exactly once in this representation and in this sense the two dimensional plane is completely covered once.  

The next step is to introduce the quantum Hilbert space, denoted ${\cal H}_q$.  This is the space of all Hilbert-Schmidt operators acting on ${\cal H}_c$ and that are generated by the non-commutative coordinates.  We denote states in ${\cal H}_c$ by $|\cdot\rangle$ and states in ${\cal H}_q$ by $|\cdot)$. The inner product on ${\cal H}_q$ is $(\phi|\psi)={\rm tr}_c(\phi^\dagger\psi)$ where ${\rm tr}_c$ denotes the trace over ${\cal H}_c$.  A general element of ${\cal H}_q$ thus has the form $|a_{n,m})=\sum_{n,m} a_{n,m} |n\rangle\langle m|$ with $\sum_{n,m} |a_{n,m}|^2<\infty$.  Note that the states $|n,m)=|n\rangle\langle m|$ form a complete orthonormal basis in ${\cal H}_q$.

From here the construction of the quantum theory proceeds as normal: one introduces observables as self-adjoint operators acting on ${\cal H}_q$ and the standard probabilistic interpretation.  To distinguish these from operators on ${\cal H}_c$, we denote them by capitals.  The only generalisation is that a position measurement must now be interpreted in the context of a weak measurement or a positive valued measure (POVM).  A detailed discussion of this can be found in \cite{scholtz}, where it was also shown that standard commutative quantum mechanics is recovered in the limit $\theta\rightarrow 0$.

The most important observable for our current purposes is the Hamiltonian given by \cite{scholtz}
\begin{equation}
H=\frac{\bar{P}{P}}{2m}+V(\hat{R}), \quad V(\hat{R})^\dagger=V(\hat{R}).
\end{equation}
Here the action of the momentum operators on a generic element  $\psi$  of ${\cal H}_q$ is defined as 
\begin{equation}
P|\psi)=|-i\hbar\sqrt{\frac{2}{\theta}}[b,\psi]),\quad \bar{P}|\psi)=|i\sqrt{\frac{2}{\theta}}\hbar[b^\dagger,\psi]).
\end{equation}
Similarly, the action of the position operators is defined as 
\begin{equation}
\hat{X}|\psi)=|\hat{x} \psi),\quad \hat{Y}|\psi)=|\hat{y} \psi),\quad \hat{R}^2=\hat{X}^2+\hat{Y}^2.
\end{equation}
Note that momentum involves a left and right multiplication, while position only involves left multiplication.  

Another useful observable is the angular momentum, which acts as follows:
\begin{equation}
L|\psi)=|\hbar[b^\dagger b,\psi])
\end{equation}
If the potential is a function of $\hat{R}$ only, this operator commutes with the Hamiltonian and is a conserved quantity.

To find the coherent state path integral action is now straightforward.  One first introduces an overcomplete set of minimal uncertainty states on ${\cal H}_c$ and ${\cal H}_q$.  For ${\cal H}_c$ they are the standard normalised Glauber coherent states
\begin{equation}
|z\rangle=e^{-|z|^2/2} e^{z b^\dagger}|0\rangle,\quad \int \frac{d\bar{z}dz}{\pi}|z\rangle\langle z|={\bf 1}_c
\end{equation}
and represent the best approximation to a position eigenstate or point in the plane.  In this sense $z$ must then be interpreted as dimensionless complex coordinates on the plane as is clear from the expectation values $x=\langle z|\hat{x}|z\rangle=\sqrt{2\theta}{\rm Re}{z}$ and $y=\langle z|\hat{y}|z\rangle=\sqrt{2\theta}{\rm Im}{z}$.  

From this, the corresponding coherent states on ${\cal H}_q$ can easily be written down
\begin{equation}
\label{csq}
|z,w)=|z\rangle\langle w|.
\end{equation}
Noting that 
\begin{equation}
|z,w)=e^{-\frac{1}{2}(\bar{z}z+\bar{w}w)}\sum_{n,m=0}^\infty \frac{z^nw^m}{\sqrt{n!m!}}|n,m),
\end{equation}
we have 
\begin{equation}
\int \frac{d\bar{z}dzd\bar{w}dw}{\pi^2}\; |z,w)(z,w|=\sum_{n,m=0}^\infty |n,m)(n,m|=\bf{1}_q.
\end{equation}

Keeping in mind that the time evolution operator acts on ${\cal H}_q$, the path integral representation of the transition amplitude in the coherent state representation (\ref{csq}) can be easily found from (\ref{action}) and is given by 
\begin{equation}
S=\int_{t_i}^{t_f}dt ( z(t),w(t)|i\hbar\frac{\partial}{\partial t}-H|z(t),w(t)).
\end{equation}
A simple computation yields the explicit form 
\begin{equation}
\label{action2dexp}
S=\int_{t_i}^{t_f}dt \left[\frac{i\hbar}{2}\left(\bar{z}\dot{z}-\dot{\bar{z}}z+\dot{\bar{w}}w-\bar{w}\dot{w}\right)-H(z,\bar{z},w,\bar{w})\right],
\end{equation}
with
\begin{equation}
H(z,\bar{z},w,\bar{w})=\frac{\hbar^2}{m\theta}\left(\left(\bar{z}-\bar{w}\right)\left(z-w\right)+1\right)+\tilde{V}(R).
\end{equation}

Here $R=\bar{z}z$ and $\tilde{V}(R)=(z,w|V(\hat{R})|z,w)={\rm tr}_c(|w\rangle\langle z|V(\hat{R})|z\rangle\langle w|)=\langle z|V(\hat{R})|z\rangle$.  Note that the function $\tilde{V}$ is different from $V$ as a normal ordering is required to replace $\hat{R}$ by its expectation value. The rest of the terms in the action are computed in a similar way, the only point of care being the right acting operators in the kinetic energy term.  The constant that appears comes from the normal ordering of right acting operators to compute the coherent state expectation value.

A more restrictive set of coherent states in which $z=w$ can also be introduced.  They satisfy an overcompleteness relation of the form 
\begin{equation}
\int \frac{d\bar{z}dz}{\pi}\; |z,z)\star (z,z|=\bf{1}_q.
\end{equation}
where $\star$ denotes the Voros product.  In \cite{scholtz} these states were used to derive the path integral action for a particle in the non-commutative plane.  To make contact with that result, we introduce a change of variables from $w$ and $z$ to $z$ and $v$ with $w=v+z$.  This gives the action
\begin{equation}
S=\int_{t_i}^{t_f}dt \left[i\hbar\left(\dot{\bar{z}}v-\bar{v}\dot{z}-\bar{v}\dot{v}\right)-\frac{\hbar^2}{m\theta}\bar{v}v-\left(\tilde{V}(R)+\frac{\hbar^2}{m\theta}\right)\right].
\end{equation}
  
Noting that this action is quadratic in $v$, the $v$ integration can be performed explicitly to yield 
\begin{equation}
S=\int_{t_i}^{t_f}dt \left[m\theta\dot{\bar{z}}\left(1+\frac{im\theta}{\hbar}\partial_t\right)^{-1}\dot{z}-\left(\tilde{V}(R)+\frac{\hbar^2}{m\theta}\right)\right],
\end{equation}
in agreement with \cite{scholtz} (note that \cite{scholtz} contains a sign misprint in the factor before $\partial_t$).

With the action in hand, we can give precise meaning to the notion of classical dynamics in the sense of a saddle point of the action.  Returning to (\ref{action2dexp}), we can easily derive the equations governing the classical dynamics:
\begin{eqnarray}
\label{eqm2d}
i\hbar\dot{\bar{w}}+\frac{\hbar^2}{m\theta}\left(\bar{z}-\bar{w}\right)&=&0,\\
-i\hbar\dot{w}+\frac{\hbar^2}{m\theta}\left(z-w\right)&=&0,\\
-i\hbar\dot{\bar{z}}-\frac{\hbar^2}{m\theta}\left(\bar{z}-\bar{w}\right)-\frac{\partial \tilde{V}}{\partial z}&=&0,\\
i\hbar\dot{z}-\frac{\hbar^2}{m\theta}\left(z-w\right)-\frac{\partial \tilde{V}}{\partial \bar{z}}&=&0.
\end{eqnarray}
Note that these equations still involve $\hbar$.  In fact, the order of the limits $\hbar\rightarrow 0$ and $\theta\rightarrow 0$ is important here.  Taking the $\theta\rightarrow 0$ limit first and then $\hbar\rightarrow 0$ gives a well defined result, while the other order does not.  In the former one of course expects, and indeed does, get the classical commutative result.

Assuming that the potential only depends on $R$, there are two constants of motion related to a $U(1)$ symmetry involving a global phase change on all variables and time translation invariance, these are the angular momentum and energy:
\begin{equation}
L=\hbar(\bar{z}z-\bar{w}w),\quad E=H(z,\bar{z},w,\bar{w})
\end{equation}
Using the equations of motion (\ref{eqm2d}) one can check explicitly that these quantities are indeed conserved.

We are not interested in the dynamics of $w$, but only the physical coordinates $z$ and would like to eliminate the former.  As the two last equations of (\ref{eqm2d}) are algebraic equation for $w$, $\bar{w}$, we can solve for them and compute the equation of motion for $z$ by substituting in the first two equations of (\ref{eqm2d}).  This yields
\begin{eqnarray}
\ddot{z}&&=-\frac{1}{m\theta}\frac{\partial \tilde{V}}{\partial \bar{z}}-\frac{i}{\hbar}\frac{d}{dt}\left(\frac{\partial \tilde{V}}{\partial \bar{z}}\right),\nonumber\\
\ddot{\bar{z}}&&=-\frac{1}{m\theta}\frac{\partial \tilde{V}}{\partial z}+\frac{i}{\hbar}\frac{d}{dt}\left(\frac{\partial \tilde{V}}{\partial z}\right).
\end{eqnarray}

We can return to the dimensionful coordinates $x$ and $y$ by writing
\begin{equation}
z=\frac{1}{2\theta}\left(x+iy\right),\quad \bar{z}=\frac{1}{2\theta}\left(x-iy\right),
\end{equation}
and
\begin{equation}
\frac{\partial}{\partial z}=\sqrt{\frac{\theta}{2}}\left(\frac{\partial}{\partial x}-i\frac{\partial}{\partial y}\right),\quad \frac{\partial}{\partial \bar{z}}=\sqrt{\frac{\theta}{2}}\left(\frac{\partial}{\partial x}+i\frac{\partial}{\partial y}\right).
\end{equation}
This gives the equation of motion 
\begin{eqnarray}
\label{eqmotion}
\ddot{x}&=&-\frac{1}{m}\frac{\partial V}{\partial x}+\frac{\theta}{\hbar}\frac{d}{dt}\left(\frac{\partial V}{\partial y}\right),\\
\ddot{y}&=&-\frac{1}{m}\frac{\partial V}{\partial y}-\frac{\theta}{\hbar}\frac{d}{dt}\left(\frac{\partial V}{\partial x}\right).
\end{eqnarray}

This is the standard Newton equations of motion, supplemented by a non-commutative correction.  In the $\theta\rightarrow 0$ limit, we recover standard Newton dynamics.  As already mentioned, the limit $\hbar\rightarrow 0$ cannot be taken before the commutative limit.  

We do not explore the consequences of this modified dynamics here, but rather postpone the in depth analysis to the three dimensional case, which is much more interesting and physically relevant.  

\section{Three dimensional fuzzy space}
\label{section 3}

In this section we study the modified classical dynamics on three dimensional fuzzy space.  The non-commutative quantum mechanics on three dimensional fuzzy space has been studied extensively in \cite{scholtz2,scholtz5,scholtz6,press}.  In these studies it was shown that this formulation reduces to commutative quantum mechanics in the commutative limit and that it is a realistic description of the physics at low energies.  At high energies there are strong deviation from commutative quantum mechanics, most notably the existence of an upper bound on the energy of a free particle \cite{scholtz5,press}, given by $E_{\rm max}=\frac{2\hbar^2}{m\lambda^2}$, and a finite density of single particle states \cite{scholtz3}.  Our interest here is to see how this translates into the classical dynamics and what observational consequences it may have.

We start by reviewing the formulation of non-commutative quantum mechanics on three dimensional fuzzy space, which follows essentially the same logic as for the two dimensional non-commutative plane.  The main difference is the modification of the coordinate algebra as the commutation relations adopted in the case of the non-commutative plane breaks rotational symmetry.  To rectify this, we adopt fuzzy sphere commutation relations
\begin{equation}
[\hat{x}_i,\hat{x}_j]=2i\lambda\varepsilon_{ijk}\hat{x}_k.
\end{equation}
Here $\lambda$ has the units of a length and $\varepsilon_{ijk}$ is the standard completely anti-symmetric tensor.  

The representation we choose for this coordinate algebra is the standard Schwinger realisation of $SU(2)$.  Thus, classical configuration space ${\cal H}_c$ is a two boson mode Fock space on which the coordinates are realised as
\begin{equation}
\hat{x}_i=\lambda a^\dagger_\alpha \sigma^{(i)}_{\alpha\beta} a_\beta.
\end{equation}
Here a summation over repeated indices is implied, $\alpha,\beta=1,2$, $\sigma^{(i)}_{\alpha\beta}$, $i=1,2,3$ are the Pauli spin matrices and $a_\alpha^\dagger$, $a_\alpha$ are standard boson creation and annihilation operators.  The radius operator is 
\begin{equation}
\hat{r}^2=\hat{x}_i\hat{x}_i =\lambda^2 \hat{n}(\hat{n}+2),
\end{equation}
with $\hat{n}=a_\alpha^\dagger a_\alpha$ the boson number operator.  Note that the radius operator is also the Casimir of $SU(2)$ and commutes with the coordinates.    As a measure of the radius we use 
\begin{equation}
\label{radiusc}
\hat{r}=\lambda (\hat{n}+1),
\end{equation}
which is to leading order in $\lambda$ the square root of $\hat{r}^2$.  Note that this representation contains each $SU(2)$ representation, and thus each quantised radius, exactly once and therefore again corresponds to a complete single covering of $R^3$, commonly referred to as fuzzy space.

The quantum Hilbert space ${\cal H}_q$ is now defined as the algebra of operators generated by the coordinates, i.e. the operators acting on ${\cal H}_c$ that commute with $\hat{r}^2$ and have a finite norm with respect to a weighted Hilbert-Schmidt inner product \cite{press}: 
\begin{eqnarray}
\label{qhilbert}
{\cal H}_q=\left\{\psi=\sum_{m_i,n_i = 0}^{\infty}C^{m_1,m_2}_{n_1,n_2} (a^\dagger_{1})^{m_1}(a^\dagger_{2})^{m_2}a_{1}^{n_1}a_{2}^{n_2}: m_1+m_2=n_1+n_2\ \ {\rm and}\ \ {\rm tr_c}(\psi^\dagger\hat{r}\psi)<\infty\right\}.
\end{eqnarray}
The inner product on ${\cal H}_q$ is
\begin{equation}
\label{innp}
	(\psi|\phi)=4\pi\lambda^2{\rm tr}_c(\psi^\dagger\hat{r}\phi)=4\pi\lambda^3{\rm tr}_c(\psi^\dagger\left(\hat{n}+1\right)\phi)
\end{equation}
with the trace taken over ${\cal H}_c$. This choice of the inner product is motivated by the observation that the norm of the operator that projects on the subspace of spheres with radius $r\leq \lambda(N+1)$, with $N$ large, corresponds to the volume of a sphere in three dimensional Euclidean space \cite{press}.

We use the standard $|\cdot\rangle$ notation for elements of ${\cal H}_c$ and $|\cdot)$ for elements of ${\cal H}_q$.  It is important to note here that, in contrast to the two dimensional non-commutative plane, the quantum Hilbert space is here restricted to only those operators on ${\cal H}_c$ that commute with the Casimir operator.  This will be an important restriction in what follows.

 Quantum observables are identified with self-adjoint operators acting on ${\cal H}_q$. We again use capitals to distinguish them from operators acting on ${\cal H}_c$.  These include the coordinates which act through left multiplication as
\begin{equation}
	\hat{X}_i|\psi)=|\hat{x}_i\psi)
\end{equation}
and the angular momentum operators which act adjointly according to
\begin{equation}
	\hat{L}_i|\psi) = |\frac{\hbar}{2\lambda}[\hat{x}_i, \psi])\quad{\rm with}\quad[\hat{L}_i,\hat{L}_j] = i\hbar\varepsilon_{ijk}\hat{L}_k.
\label{eq:nc-angular-ops}
\end{equation}
The non-commutative analogue of the Laplacian is defined as
\begin{equation}
 \hat{\Delta}|\psi)=-|\frac{1}{\lambda\hat{r}}[\hat{a}^\dagger_\alpha,[\hat{a}_\alpha,\psi]])=|\frac{1}{\lambda^2\left(\hat{n}+1\right)}[\hat{a}^\dagger_\alpha,[\hat{a}_\alpha,\psi]])
\label{eq:nc-laplacian}
\end{equation}
and can be shown to commute with the three angular momentum operators \cite{press}.  

The Hamiltonian is given by
\begin{equation}
	\hat{H}=-\frac{\hbar^2}{2m}\hat{\Delta}+V(\hat{R})
\end{equation}
with $\hat{R}$ the radius operator that acts as 
\begin{equation}
\label{radiusq}
\hat{R}|\psi)=|\lambda(\hat{n}+1)\psi),\, \hat{n}=a^\dagger_\alpha a_\alpha.
\end{equation}

From the discussion above it should be clear that the angular momentum operators commute with the Hamiltonian and are therefore conserved.  There is a further important conserved quantity, namely, the operator $\hat\Gamma$, which acts as follows
\begin{equation}
\hat\Gamma|\psi)=|[a^\dagger_\alpha a_\alpha,\psi]).
\end{equation}
It is simple to check explicitly that it does in fact commute with the Hamiltonian.  

To facilitate the construction of the classical dynamics on fuzzy space, we enlarge the quantum Hilbert space ${\cal H}_q$ to include all Hilbert-Schmidt operators acting on ${\cal H}_c$, i.e. all operators with finite norm generated by the creation and annilhilation operators $a^\dagger_\alpha$ and $a_\alpha$.  The inner product is still given by (\ref{innp}).  We denote this enlarged space by ${\cal H}^0_q$.  Clearly ${\cal H}_q\subset {\cal H}^0_q$.  From the definition of ${\cal H}_q$, it is then clear that physical states, i.e. states that belong to the subspace ${\cal H}_q$ must satisfy the constraint
\begin{equation}
\label{constr}
\hat\Gamma|\psi)=0.
\end{equation}
Note that since $\hat\Gamma$ is conserved, initial states that satisfy this condition, will do so at all times. Below we use this property explicitly in the construction of the path integral representation of physical transition amplitudes.

We now proceed with the construction of the path integral representation of physical transition amplitudes.  The first step is to get rid of the weighted inner product in (\ref{innp}).  This can be done by redefining the wave functions as follows
\begin{equation}
\label{rwf}
\tilde{\psi}=\sqrt{\hat{r}}\psi.
\end{equation}
The inner product then assumes the standard form 
\begin{equation}
\label{innp1}
	(\tilde{\psi}|\tilde{\phi})=4\pi\lambda^2{\rm tr}_c(\tilde{\psi}^\dagger\tilde{\phi})
\end{equation}
Upon doing this, we must, however, also transform the Hamiltonian, or any other observable, as follows
\begin{equation}
\label{trans}
\hat{\tilde{H}}=\sqrt{\hat{r}} H\frac{1}{\sqrt{\hat{r}}}.
\end{equation}
From here on we work with this quantum Hilbert space in which the inner product is given by (\ref{innp1}) and observables are transformed as in (\ref{trans}).  We denote this space by $\tilde{{\cal H}}_q$ and its enlargement by $\tilde{{\cal H}}^0_q$.  Note that the constants of motion $\hat\Gamma$ and $\hat{L}_i$ are unchanged by this transformation.

It is obvious that $\hat\Gamma$ and $\hat{L}_i$ also commute with $\hat{\tilde{H}}$ and are conserved under the time evolution generated by this Hamiltonian.  It is also clear that physical states are still characterised by the constraint
\begin{equation}
\hat\Gamma|\tilde{\psi})=0.
\end{equation}

We introduce the standard minimum uncertainty states on ${\cal H}_c$ as Glauber coherent states, which form an overcomplete basis
\begin{eqnarray}
&&|z_\alpha\rangle=e^{-|\bar{z}_\alpha z_\alpha/2}e^{z_\alpha a^\dagger_\alpha}|0\rangle,\nonumber\\
&&\int \frac{d\bar{z}_\alpha dz_\alpha}{\pi^2} |z_\alpha\rangle\langle z_\alpha|=\bf{1}_c.
\end{eqnarray}
The dimensionful physical coordinates are now identified as
\begin{equation}
\label{coord}
x_i=\langle z_\alpha |\hat{x}_i| z_\alpha\rangle=\lambda \bar{z}_\alpha \sigma^{(i)}_{\alpha\beta} z_\beta.
\end{equation}

As in the two dimensional non-commutative plane, we can correspondingly introduce coherent states on $\tilde{{\cal H}}^0_q$ as
\begin{equation}
|z_\alpha, w_\alpha)=|z_\alpha\rangle\langle w_\alpha|.
\end{equation}
They are overcomplete and 
\begin{equation}
\int \frac{d\bar{z}_\alpha dz_\alpha d\bar{w}_\alpha dw_\alpha}{\pi^4} |z_\alpha,z_\alpha)(z_\alpha, w_\alpha|=\bf{\tilde{1}}^0_q.
\end{equation}

It is important to note that the states $|z_\alpha, w_\alpha)$ are not all physical.  However, we are interested in physical transition amplitudes, which implies that if the initial state is physical, all the states at intermediate times are also physical as $\hat\Gamma$ commutes with $\hat{\tilde{H}}$.  As the states $|z_\alpha, w_\alpha)$ resolve the identity on $\tilde{{\cal H}}^0_q$, we can safely use them to insert the identity at intermediate times in a time slicing procedure, provided that the initial state is physical.  Indeed, if this is done, the constraint must appear as a conserved quantity in the resulting action and we must simply require it to vanish to satisfy the condition of physicality of the initial state.  

Following this approach, the general result of (\ref{action}) is still applicable and to obtain the path integral action we therefore only have to compute the action 
\begin{equation}
\label{temp}
S=\int_{t_i}^{t_f}dt ( z_\alpha(t),w_\alpha(t)|i\hbar\frac{\partial}{\partial t}-\hat{\tilde{H}}|z_\alpha(t),w_\alpha(t)).
\end{equation}

To simplify matters, it is convenient to introduce dimensionless quantities from here on.  We reserve capitals to denote these.  Introduce the following time scale, $t_0$, energy scale, $e_0$, dimensionless time, $T$, dimensionless coordinates, $X_i$, and dimensionless energy, $E$,
\begin{equation}
\label{dim}
t_0=\frac{m\lambda^2}{\hbar},\quad e_0=\frac{\hbar}{t_0},\quad T=\frac{t}{t_0},\quad X_i=\frac{x_i}{\lambda},\quad E=\frac{e}{e_0}.
\end{equation}
The dimensionless action $\tilde{S}=\frac{S}{\hbar}$ can than be explicitly computed.  The computation is slightly more involved than in the case of the two dimensional non-commutative plane, but still straightforward.  We find
\begin{equation}
\label{action3d}
\tilde{S}=\int_{T_i}^{T_f}dT \left[\frac{i}{2}\left(\bar{z}_\alpha\dot{z}_\alpha-\dot{\bar{z}}_\alpha z_\alpha+\dot{\bar{w}}_\alpha w_\alpha-\bar{w}_\alpha\dot{w}_\alpha\right)-\tilde{H}(z_\alpha,\bar{z}_\alpha,w_\alpha,\bar{w}_\alpha)\right],
\end{equation}
where
\begin{equation}
\label{action3dh}
\tilde{H}(z,\bar{z},w,\bar{w})=\left(f_1(R)\bar{z}_\alpha z_\alpha-f_2(R)\left(\bar{z}_\alpha w_\alpha+z_\alpha \bar{w}_\alpha\right)+f_3(R)\bar{w}_\alpha w_\alpha\right)+W(R).
\end{equation}
Here 
\begin{eqnarray}
R&&=\bar{z}_\alpha z_\alpha,\label{a}\\
f_1(R)&&=\frac{1}{2}\langle z_\alpha|\frac{1}{\hat{n}+2}|z_\alpha\rangle,\label{b}\\
f_2(R)&&=\frac{1}{2}\langle z_\alpha|\frac{1}{\sqrt{(\hat{n}+1)(\hat{n}+2)}}|z_\alpha\rangle,\label{c}\\
f_3(R)&&=\frac{1}{2}\langle z_\alpha|\frac{1}{\hat{n}+1}|z_\alpha\rangle,\label{d}\\
W(R)&&=\frac{1}{e_0}\langle z_\alpha|V(\hat{R})|z_\alpha\rangle+2f_3(R)\equiv\tilde{V}(R)+2f_3(R).\label{e}
\end{eqnarray}
Note that $R$, all the $f_i(R)$ and $W(R)$ are dimensionless.   

The equations of motion determining the classical dynamics can now be easily derived and are given by
\begin{eqnarray}
\dot{z}_\alpha&=&-i\frac{\partial \tilde{H}}{\partial \bar{z}_\alpha},\label{eqma}\\
\dot{\bar{z}}_\alpha&=&i\frac{\partial \tilde{H}}{\partial z_\alpha},\label{eqmb}\\
\dot{w}_\alpha&=&i\frac{\partial \tilde{H}}{\partial \bar{w}_\alpha},\label{eqmc}\\
\dot{\bar{w}}_\alpha&=&-i\frac{\partial \tilde{H}}{\partial \bar{w}_\alpha}.\label{eqmd}
\end{eqnarray}
There are five conserved quantities, four related to a $U(2)$ symmetry and the fifth a conserved energy related to time translation invariance. These are easily found to be
\begin{eqnarray}
\label{con3d}
\Gamma&=&\bar{z}_\alpha z_\alpha-\bar{w}_\alpha w_\alpha,\nonumber\\
L_i&= &\bar{z}_\alpha \sigma^{(i)}_{\alpha\beta} z_\beta-\bar{w}_\alpha \sigma^{(i)}_{\alpha\beta} w_\beta,\nonumber\\
E&=&\tilde{H}(z,\bar{z},w,\bar{w}).
\end{eqnarray}
It can be checked directly from (\ref{eqma})-(\ref{eqmd}) that these quantities are indeed constant in time.  The first, $\Gamma$, is simply the expectation value of the conserved quantity $\hat{\Gamma}$ in the state $|z_\alpha, w_\alpha)$ and therefore naturally conserved.  This quantity also determines whether states are physical or not and must vanish for physical states.  We must therefore require $\Gamma=0$.  The $L_i$ are just the expectation values of the momentum operators $\hat{L}_i$ in the same state and therefore also conserved.  Finally,  $\tilde{H}(z,\bar{z},w,\bar{w})$ is just the Hamiltonian and, as it is not explicitly time dependent, conserved.

Our interest is not in the equations of motion of the $z_\alpha$ and $w_\alpha$, but rather in the equations of motion of the physical, dimensionless coordinates $X_i=\frac{x_i}{\lambda}$ with the $x_i$ given in (\ref{coord}).  We must therefore eliminate $z_\alpha$ and $w_\alpha$ in favour of these.  This is a long and tedious calculation that can fortunately be done efficiently with Mathematica.  The easiest way to proceed is to first paramaterise the $z_\alpha$ as follows:
\begin{eqnarray}
z_1&=&\sqrt{R}\cos(\frac{\theta}{2})e^{-i\frac{\phi}{2}}e^{i\gamma},\nonumber\\
z_2&=&\sqrt{R}\sin(\frac{\theta}{2})e^{i\frac{\phi}{2}}e^{i\gamma},
\end{eqnarray}
and the corresponding complex conjugates where $R>0$, $\theta$, $\phi$ and $\gamma$ real.  With this parameterisation the coordinates take the standard form in spherical coordinates
\begin{eqnarray}
X_1&=&R\sin\theta\cos\phi,\nonumber\\
X_2&=&R\sin\theta\sin\phi,\nonumber\\
X_3&=&R\cos\theta.
\end{eqnarray}
Note that the global phase $\gamma$ drops out from these expressions, but not from the time derivatives.

One now proceeds as follows: Solve for the $w_\alpha$ from the algebraic equations (\ref{eqma}), (\ref{eqmb}) in terms of the $z_\alpha$ and their time derivatives.  These expressions also contain $\dot{\gamma}$.  Solve $\dot{\gamma}$ from the constraint $\Gamma=0$.  Substitute this back into the expressions for the second order time derivatives of the coordinates, computed using the equations of motion (\ref{eqma})-(\ref{eqmd}).  Although the intermediate steps are involved, the final result is fairly simple and reads as follows:
\begin{equation}
\label{eqmf}
\ddot{\vec{X}}_\pm=a_\pm(R,V) \vec{X}+b_\pm(R,V)\left(\vec{X}\times\dot{\vec{X}}\right)+c_\pm(R)\left(\left(\vec{X}\times\dot{\vec{X}}\right)\times\dot{\vec{X}}\right).
\end{equation}
Here $R^2=\vec{X}\cdot\vec{X}$ and 
\begin{eqnarray}
\label{eqmfa}
a_\pm(R,V)&=&4R^2f_2(R)^2g_1(R)\pm\frac{g_2^\prime(R)}{R}\sqrt{4R^2f_2(R)^2-\dot{\vec{X}}\cdot\dot{\vec{X}}},\nonumber\\
b_\pm(R,V)&=&\frac{g_2^\prime(R)}{R}\pm g_1(R)\sqrt{4R^2f_2(R)^2-\dot{\vec{X}}\cdot\dot{\vec{X}}},\nonumber\\
c_\pm(R)&=&g_1(R).
\end{eqnarray}
Here
\begin{eqnarray}
g_1(R)&=&\frac{1}{R^2}+\frac{f_2^\prime(R)}{f_2(R)R},\nonumber\\
g_2(R)&=&R\left(f_1(R)+f_3(R)\right)+W(R)
\end{eqnarray}
and the prime denotes derivation with respect to $R$.

The dimensionless conserved quantities can also be computed, but now there are only four as the constraint $\Gamma=0$ is satisfied by construction.  They are
\begin{eqnarray}
\label{const}
\vec{L}_\pm&=&\frac{1}{4f_2(R)^2R^2}\left[\sqrt{4R^2f_2(R)^2-\dot{\vec{X}}\cdot\dot{\vec{X}}}\left(\vec{X}\times\dot{\vec{X}}\right)\pm \left(\vec{X}\times\dot{\vec{X}}\right)\times\dot{\vec{X}}\right],\label{consta}\\
E_\pm&=&g_2(R)\pm \sqrt{4R^2f_2(R)^2-\dot{\vec{X}}\cdot\dot{\vec{X}}}.\label{constb}
\end{eqnarray}
Note that there are two branches denoted $\pm$.  Indeed, it is clear from (\ref{constb}) that the branch is determined by the sign of $E-g_2(R)$.  The use of two branches in the equations of motion is inconvenient, but it turns out that a unified treatment is possible when one considers the radial motion in terms of an effective potential.  We return to this in the next section.

One can benchmark these results in a number of ways.  Firstly, one can check, using the equations of motion (\ref{eqmf}), that the constants of motion are indeed constant in time, which turns out to be the case.  Secondly, one can solve the equations of motion (\ref{eqma})-(\ref{eqmd}) numerically and check that this also solves (\ref{eqmf}).  This also checks out.  In this process one also finds that both branches are needed to describe the full dynamics. We discuss these equations of motion in more detail in the next section.

This is the most general form of the equations of motion. Indeed, in this form one may view the $f_i(R)$ as arbitrary functions, but note that if this is done there is a redundancy in $f_1(R)$, $f_3(R)$ and $W(R)$ as only the combination of $g_2(R)$ plays a role.  Since this turns out to be a useful point of view, we explore it further in section \ref{geneqm}.  For our current purposes though, we continue to compute the functions $f_i(R)$ as they appear in (\ref{b})-(\ref{d}). 

To do this, we note that the coherent state (summation over repeated indices is implied) $|z_\alpha\rangle=e^{-\frac{\bar{z}_\alpha z_\alpha}{2}} e^{z_\beta a^\dagger_\beta }|0\rangle$ can be rewritten, upon introducing a new creation operator $A^\dagger=\frac{1}{\sqrt{R}}z_\beta a^\dagger_\beta$ ($R=\bar{z}_\alpha z_\alpha$), as  $|z_\alpha\rangle=e^{-\frac{R}{2}}e^{R A^\dagger} |0\rangle$.  It then follows easily that for any function $g(\hat{n}+1)$
\begin{equation}
\label{g}
g(R)\equiv\langle z_\alpha|g(\hat{n}+1)|z_\alpha\rangle=e^{-R}\sum_{n=0}^\infty g(n+1)\frac{R^n}{n!}.
\end{equation}
From this we also easily deduce the general relation
\begin{equation}
\label{derf}
\langle z_\alpha|g(\hat{n}+2)|z_\alpha\rangle=g(R)+\frac{dg(R)}{dR}.
\end{equation}
Similar relations can be derived for $g(\hat{n}+k)$, for $k$ a positive positive integer, by iterating (\ref{derf}).

By explicit summation, we can now easily compute $f_3(R)$ exactly.  Using (\ref{derf}), we can extract $f_1(R)$ exactly.  Finally, upon noting 
\begin{equation}
\langle z_\alpha|\frac{1}{(\hat{n}+1)^k}|z_\alpha\rangle=e^{-R}\sum_{n=0}^\infty \frac{1}{(n+1)^k}\frac{R^n}{n!}\sim\frac{1}{R^k}
\end{equation}
for large $R$, we can extract the large $R$ behaviour of $f_2$ through an expansion in orders of $\frac{1}{\hat{n}+1}$.  The final result is
\begin{eqnarray}
\label{first}
f_1(R)&=&\frac{1}{2R}-\frac{1-e^{-R}}{2R^2}\approx\frac{1}{2R}-\frac{1}{2R^2},\nonumber\\
f_2(R)&\approx&\frac{1}{2R}-\frac{1}{4R^2}-\frac{1}{16R^3},\nonumber\\
f_3(R)&=&\frac{1-e^{-R}}{2R}\approx\frac{1}{2R},
\end{eqnarray}

When one is interested in long length scales, it is sufficient to approximate these functions by
\begin{equation}
\label{zero}
f_i(R)=\frac{1}{2R},\;\forall i.
\end{equation}

In the lowest order approximation (\ref{zero}) the equations and constant of motion simplify considerably and provide a useful benchmark for understanding the dynamics.  Let us therefore consider this approximation.  Substituting (\ref{zero}) in (\ref{eqmf}) and (\ref{eqmfa}) yields
\begin{equation}
\label{eqmflong}
\ddot{\vec{X}}_\pm=\frac{W'(R)}{R}\left[\left(\vec{X}\times\dot{\vec{X}}\right)\pm\sqrt{1-\dot{\vec{X}}\cdot\dot{\vec{X}}}\;\vec{X}\right],
\end{equation}
The dimensionless conserved quantities are
\begin{eqnarray}
\label{constlong}
\vec{L}_\pm&=&\sqrt{1-\dot{\vec{X}}\cdot\dot{\vec{X}}}\left(\vec{X}\times\dot{\vec{X}}\right)\pm\left(\left(\vec{X}\times\dot{\vec{X}}\right)\times\dot{\vec{X}}\right),\label{constlonga}\\
E_\pm&=&1\pm\sqrt{1-\dot{\vec{X}}\cdot\dot{\vec{X}}}+W(R).\label{constlongb}
\end{eqnarray}

Equations (\ref{eqmf})  and (\ref{eqmflong}) have rather interesting consequences as they suggest that the dimensionless speed $V^2=\dot{\vec{X}}\cdot\dot{\vec{X}}$ of a test particle is limited in a generally spatial dependent way determined by the function $f_2(R)$.  At long length scales ($R>>1$) it becomes spatially independent and $V^2\leq 1$.   Related to this, the dimensionless kinetic energy $E_k=1\pm\sqrt{1-\dot{\vec{X}}\cdot\dot{\vec{X}}}$ is bounded by $E_k\leq 2$.   Note that energies $E_k>1$ are described by the plus branch.  This bound on the energy is in complete agreement with the bound found on the quantum level \cite{scholtz5,press}.  The bound on the speed of an object comes as a surprise and closer scrutiny traces it back to the condition of physicality (\ref{constr}) of the wave functions.  More insight can be obtained by considering the dimensionful form of the equations of motion (\ref{eqmflong})
\begin{equation}
\label{eqmfd}
\ddot{\vec{x}}_\pm=\frac{w'(r)}{mr}\left[\frac{m\lambda}{\hbar}\left(\vec{x}\times\dot{\vec{x}}\right)\pm\sqrt{1-\left(\frac{m\lambda}{\hbar}\right)^2\dot{\vec{x}}\cdot\dot{\vec{x}}}\;\vec{x}\right],
\end{equation}
and conserved quantities
\begin{eqnarray}
\label{constmd}
\vec{\ell}_\pm&=&\hbar\vec{L}=m\left[\sqrt{1-\left(\frac{m\lambda}{\hbar}\right)^2\dot{\vec{x}}\cdot\dot{\vec{x}}}\;\left(\vec{x}\times\vec{\dot{x}}\right)\pm \frac{m\lambda}{\hbar}\left(\left(\vec{x}\times\dot{\vec{x}}\right)\times\dot{\vec{x}}\right)\right],\\
e_\pm&=&\frac{\hbar^2}{m\lambda^2}\left[1\pm\sqrt{1-\left(\frac{m\lambda}{\hbar}\right)^2\dot{\vec{x}}\cdot\dot{\vec{x}}}\;\;\right]+w(r).
\end{eqnarray}
Here $w(r)=V(r)+\frac{2\hbar^2 f_3(r/\lambda)}{m\lambda^2}$ with $V(r)$ the dimensionful potential. In this form one recognises the bound on the speed of a test particle as $v_0=\frac{\hbar}{m\lambda}$. Also note that upon restoration of dimensions, the dimensionful length scale at which the limiting speed becomes spatially independent is the non-commutative parameter $\lambda$.  As this is presumably a very short scale, the limiting speed can essentially be viewed as spatially independent.  However, if the function $f_2(R)$ is treated more generally as described in section \ref{geneqm}, this may not be the case.   We postpone further discussion of this result to section \ref{discussion} after we have  established some other results that have a bearing on this.

Finally, note that, apart from a singular correction in the potential $w$, which vanishes in the $\hbar\rightarrow 0$ limit, one recovers the standard Newton equations in the $\lambda\rightarrow 0$ limit for the minus branch.  This again emphasises that the commutative and classical limits have to be taken with great care as the order matters.  In fact, here it is not possible to take either of these limits without encountering a singularity, the only sensible limit seems to be one in which the ratio between $\hbar$ and $\lambda$ is kept fixed.  In section \ref{discussion}, we argue that this is in fact also necessary from other physical considerations.

\subsection{General properties of orbitals}
\label{genorb}

The first step in understanding the motion implied by (\ref{eqmf}) and (\ref{eqmflong}) is to understand the relationship between the different conserved quantities, velocities and acceleration.  For simplicity we consider the dimensionless quantities.  It is straightforward to establish the following relations that hold on both branches and for the general equation of motion (\ref{eqmf}) and their long scale approximation (\ref{eqmflong}):
\begin{eqnarray}
\label{rel}
\vec{L}_\pm\cdot\dot{\vec{X}}&=&0,\label{rela}\\
\vec{L}_\pm\cdot\ddot{\vec{X}}_\pm&=&0,\label{relb}\\
\vec{L}_\pm\cdot\vec{X}&=&\mp\vec{L}\cdot\vec{L}\equiv \mp L^2\label{relc}
\end{eqnarray}

We note that $\vec{L}\cdot\vec{X}$ is conserved in time.  Note that this result contrasts with standard central potential motion for which $\vec{L}\cdot\vec{X}=0$.  The motion is, however, still planar as in the case of a standard central potential, but the plane is displaced along the direction of $\vec{L}$, leading to  $\vec{L}\cdot\vec{X}\ne 0$.  Specifically in the case of gravity this implies that the mass creating the gravitational force no longer lies in the plane of motion.   

Another important point to note is that the conserved quantity $\vec{X}\cdot\vec{L}$ switches signs between the two branches.  This means that the dynamics of the two branches do not mix, except in the case when $L=0$.  The minus branch reduces to standard Newton dynamics in the commutative limit and has the kinetic energy $E_k<1$.  Note that this also brings about an asymmetry: The plane of motion is always displaced in the direction of $\vec{L}$ for the minus branch and oppositely for the plus branch.  

Finally, it is convenient to introduce the vector $\vec{X}^\star=\vec{X}-\vec{L}$, which describes the motion in the plane and to note that $\dot{\vec{X}}=\dot{\vec{X}}^\star$.
 
\begin{figure}[t]
    \centering
   \begin{tabular}{c c}
    (a)&(b)\\
  \includegraphics[width=0.5\textwidth]{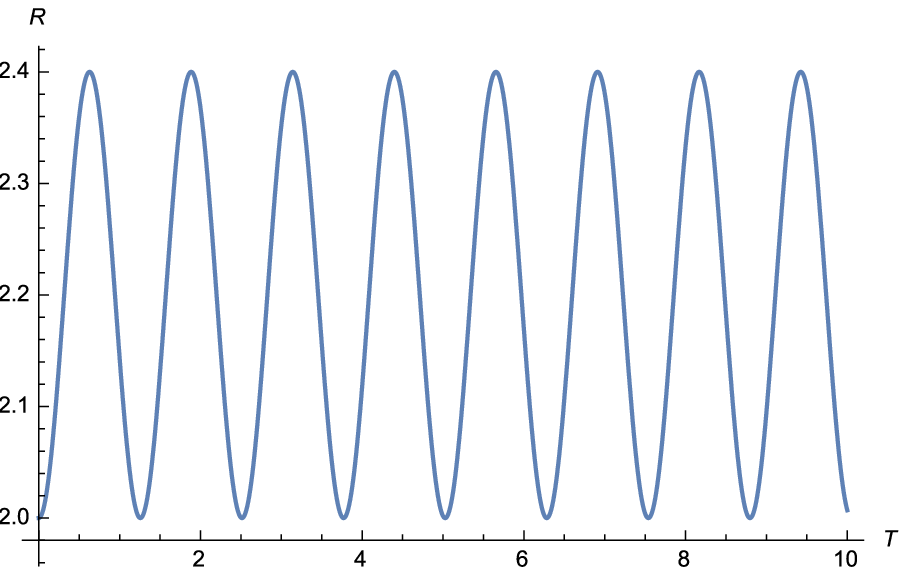}$\quad$&\includegraphics[width=0.5\textwidth]{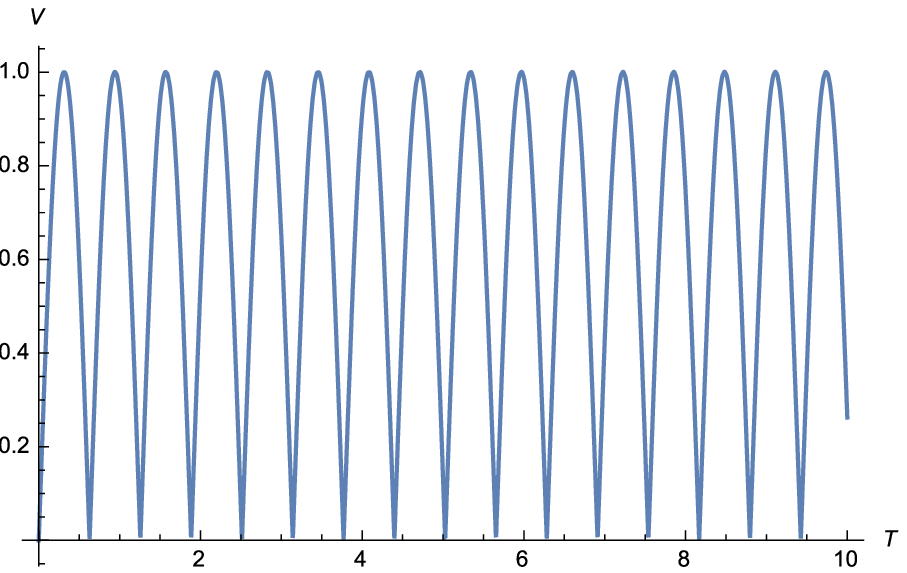}
	\end{tabular}
  \caption{Radius (a) and speed (b) of a particle subjected to a constant force in the outwards radial direction.}
  \label{figcf}
\end{figure}

The dynamics implied by (\ref{eqmf}) and (\ref{eqmflong}) is quite counter intuitive and it is useful to first develop some feeling for its content.   One of the outstanding features of these equations of motion is the appearance of a limiting speed.  One of the obvious question is what happens if an object is accelerated up to this limiting speed?  To develop some understanding of this, we focus on the long scale approximation  (\ref{eqmflong}).  Let us therefore consider these equations of motion in the presence of a constant outwards radial force, i.e., a potential of the form $-\beta R$, $\beta>0$.  As a benchmark, we first integrate the equations of motion (\ref{eqma})-(\ref{eqmd}) with this potential.  We take $\beta=5$ and as initial conditions $z_1=\bar{z}_1=w_1=\bar{w}_1=1$ and compute $w_1,\bar{w}_1$ from the constraint for physicality of the wave function in (\ref{con3d}).  For the coordinates, this choice corresponds to the initial conditions $\vec{X}=\{2,0,0\}$ and $\vec{V}=\{0,0,0\}$.  The results are shown in figures \ref{figcf} (a), (b).  

The surprise is that the motion is not simply a constant accelerated radially outward motion as one would naively expect.  The test particle accelerates until it reaches the limiting speed $V=1$, then it starts to de-accelerate.  When its speed vanishes, its radial motion reverses and it continues accelerating radially inwards until it reaches the limiting speed after which it again starts to de-accelerate until it speed vanishes and the cycle is repeated.  Exactly the same result is obtained by integrating (\ref{eqmflong}), but in this case one has to switch between the branches at the turning points in the speed (the change in sign of the acceleration is related to the flip in sign between the two branches).  To understand the origin of this oscillatory motion, we return to the conserved energy and compute the effective potential for the radial motion.  We therefore set $W(R)=-\beta R$ (we ignore the non-commutative correction to the potential here):

\begin{equation}
E_\pm=1\pm\sqrt{1-\dot{\vec{X}}\cdot\dot{\vec{X}}}-\beta R.
\end{equation}
Using $\dot{\vec{X}}^2=\dot{R}^2+\frac{L^2}{R^2}$, which can easily be checked from (\ref{constlonga}), this can be rewritten as 
\begin{equation}
\label{effplongcf}
\dot{R}^2+2(E-1)\beta R+\beta^2 R^2+\frac{L^2}{R^2}\equiv \dot{R}^2+V_{\rm eff}=E(2-E).
\end{equation}
We observe that this is indeed a harmonic oscillator potential with shifted minimum and a repulsive barrier at the origin.  Note that for $E> 2$, the left hand is positive, but the right hand is negative and that this equation cannot be satisfied for these energies.  This again demonstrates the cut-off in energy at $E=2$ referred to earlier.  

\begin{figure}[t]
   \centering
   \includegraphics[width=0.8\textwidth]{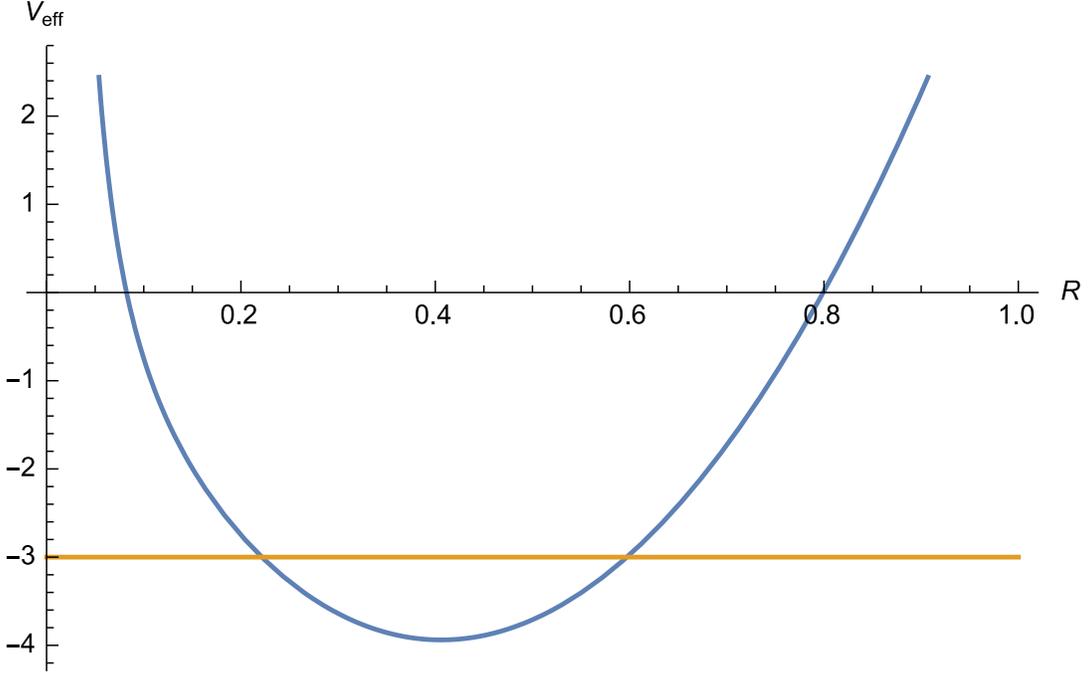}
     \caption{Effective potential for the radial motion in the presence of a radially outwards constant force. In this figure $\beta=5$, $E=-1$ and $L=0.1$}
  \label{effpcf}
 \end{figure}

In the case above, $L=0$ and the repulsive barrier is absent.  In general for $L\neq 0$, it is present and the generic effective potential is shown in figure \ref{effpcf} with $\beta=5$, $E=-1$ and $L=0.1$.  Also shown is the right hand of (\ref{effplongcf}) (horizontal line).  The points where the horizontal line cuts the curve of $V_{\rm eff}$ are the turning points of the radial motion as $\dot R=0$ at these points.  The origin of the oscillatory motion seen in figure (\ref{figcf}) should now be clear.  Also note that the analysis in terms of the effective potential is independent of the branch.  The message to take away from this exercise is that the square root based dispersion relation in (\ref{constlongb}), which is also the source of the limiting speed, can give rise to rather peculiar and counterintuitive dynamical behaviour.  However, when reformulated in terms of an effective potential, the dynamical behaviour becomes very transparent.

We now turn to the case of gravity for which the dimensionless potential reads
\begin{equation}
\label{gravpot}
W(R)=-\frac{\beta}{R}, \quad\beta=\frac{GMm^2\lambda}{\hbar^2}-1
\end{equation}

For our present purposes it is again sufficient to consider only the long length scale behaviour where we can approximate the functions $f_i(R)$ as in (\ref{zero}).   We again construct the effective potential, which now reads
\begin{equation}
\label{effplong}
\dot{R}^2+\frac{2(E-1)\beta}{R}+\frac{\beta^2+L^2}{R^2}\equiv \dot{R}^2+V_{\rm eff}=E(2-E).
\end{equation}
Firstly, note that the $1/R$ term in the effective potential (\ref{effplong}) switches sign between $E<1$ and $E>1$ and that the effective potential is strictly repulsive for $E>1$.  This is again a manifestation of the two branches already mentioned.  Secondly, note that for $E>2$, the left hand side of (\ref{effplong}) is strictly positive, while the right hand side is negative, resulting in the energy cut-off $E<2$ observed before.  Thirdly, note that there is always a repulsive barrier, even when $L=0$ on the short to medium length scales.  This is quite different to the commutative case where the centrifugal term stabilises the orbits.  It should, however, be kept in mind that the behaviour of the effective potential at short length scales may be drastically altered by the short length scale corrections to the functions $f_i(R)$.  
 
\begin{figure}[t]
    \centering
   \begin{tabular}{c c c}
    (a)&(b)&(c)\\
  \includegraphics[width=0.3\textwidth]{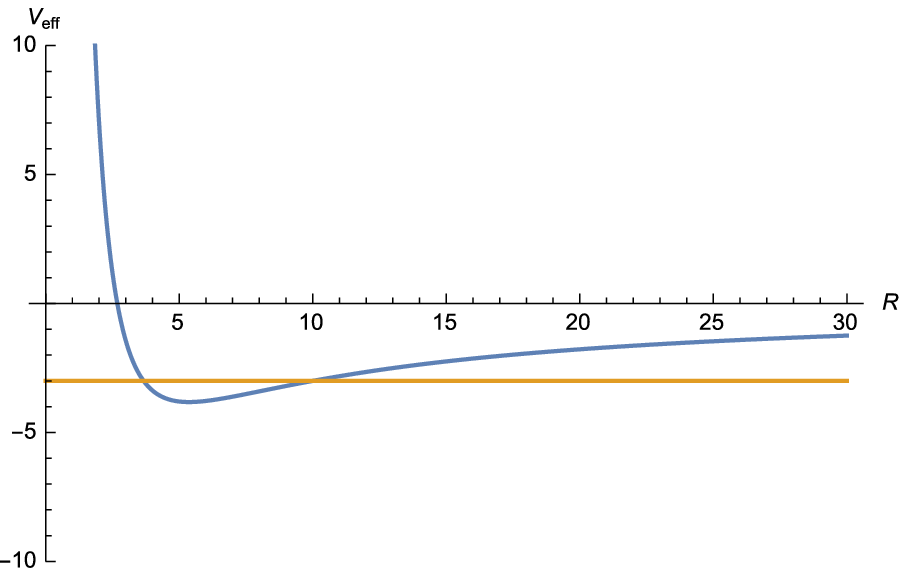}$\quad$&\includegraphics[width=0.3\textwidth]{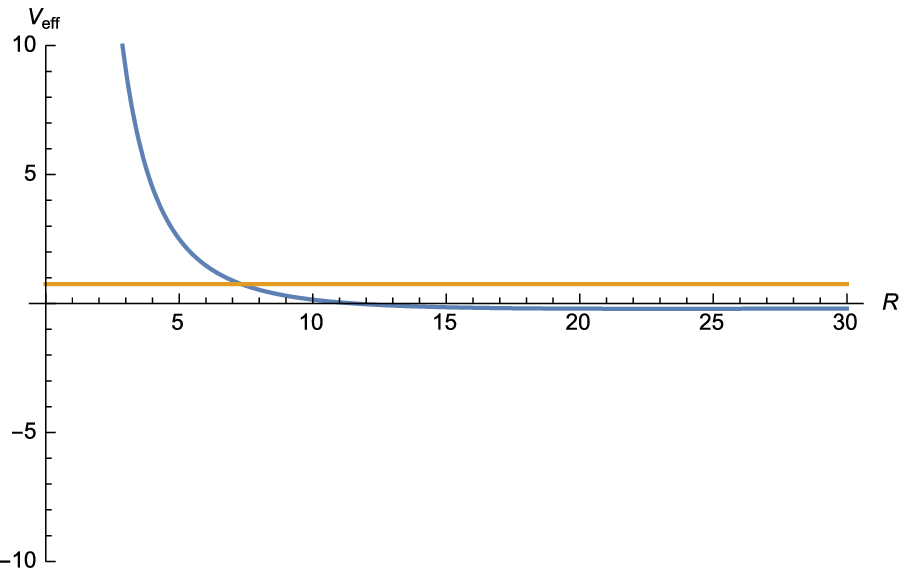}&\includegraphics[width=0.3\textwidth]{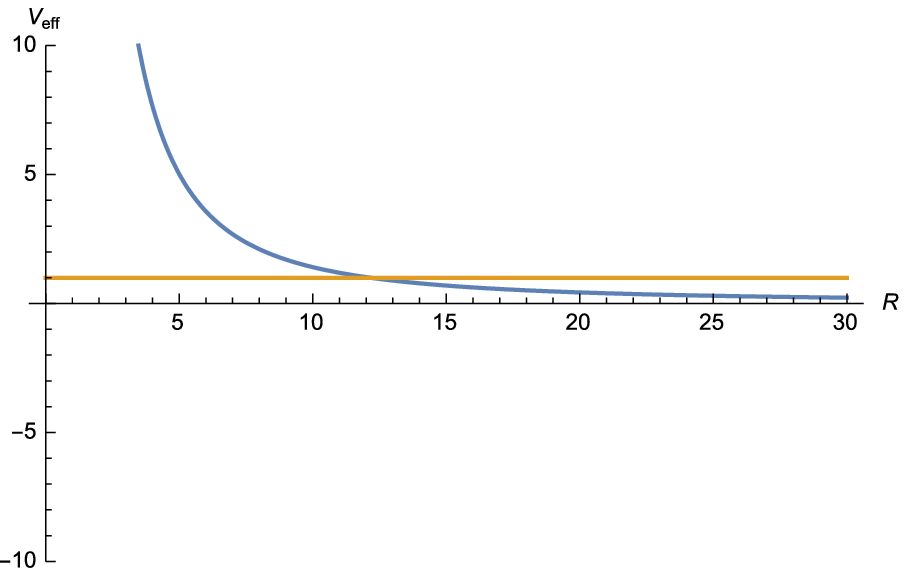}
	\end{tabular}
  \caption{Turning points of the orbitals for $E<0$ ($E=-0.5$) in (a), $0<E<1$ ($E=0.5$) in (b) and $1<E<2$ ($E=1.1$) in (c). The horizontal lines are the values of $E(2-E)$. $\beta=10$ and $L=0.1$ in these plots.}
  \label{turn}
\end{figure}

Since $\dot{\vec{X}}=\dot{\vec{X}}^\star$ the points where $E(2-E)-V_{\rm eff}$ vanishes are the turning points of the orbitals where $\dot{R}=\dot{R}^\star=0$ (see also (\ref{rad})).  These are the points where the line $E(2-E)$ intersects the curve of $V_{\rm eff}$ in figure \ref{turn}.  This is shown in figure \ref{turn} for $E<0$ (a), $0<E<1$ (b) and $1<E<2$ (c).  When $E<0$ as shown in figure \ref{turn}(a) there are two turning points, where $R$ reaches its maximum and minimum.  The motion is elliptic and the turning points represent the closest and furthest points of the orbit.  When $0<E<1$ as in figure \ref{turn}(b), there is only one turning point, despite the fact that the potential still has an attractive tail.  The particle is unbound and escapes to infinity.  When $1<E<2$ as in figure \ref{turn}(c), the potential is repulsive and a particle placed anywhere accelerates to infinity. Note, though, that its energy and speed is bound by 2 and 1, respectively.  When the turning points coincide, which only happens at the minimum of the potential, the motion is circular.  For this to happen $E$ and $L$ must be related in a specific way.
 
\subsection{Precession in a gravitational potential}

One expects that the modified dynamics implied by the non-commutativity can cause precession of elliptic orbitals and we investigate this possibility here.  It turns out that this is only possible if the short length scale corrections in the functions $f_i(R)$ are included.   Since we are interested in bound orbitals, we take $E<0$ from here on.

We start by deriving the general expression for the precession angle.  Without loss of generality we can choose $\vec{L}$ along the z-direction and for our present purpose it is also convenient to restore dimensions from here on.  Let us introduce the vector $\vec{x}^\star=\vec{x}-\frac{\lambda}{\hbar}\vec{\ell}$.  From (\ref{relc}) it has the property $\vec{\ell}\cdot\vec{x}^\star=0$ and thus represents the rotating vector in the plane of motion.  This vector only depends on the azimuthal angle $\phi$ and not on $\theta$, as is the case with $\vec{x}$.  To establish precession, we must therefore compute the dependence of this vector on $\phi$.  

To do this, we note from (\ref{relc}) and our choice of $\vec{\ell}$ along the z-axis ($\hat{\ell}$ denotes the unit vector and $\ell^2=\vec{\ell}\cdot\vec{\ell}$):
\begin{equation}
\hat{\ell}\cdot\vec{x}=r\cos\theta=\frac{\lambda}{\hbar}\ell,\label{press1}
\end{equation}
Introducing $\vec{x}\cdot\vec{x}=r^2$ and $\vec{x}^\star\cdot\vec{x}^\star={r^\star}^2$, we also have
\begin{equation}
\label{rrstar}
r^2={r^\star}^2+\frac{\lambda^2}{\hbar^2}\ell^2.
\end{equation}
From this we obtain
\begin{eqnarray}
\label{theta}
\cos\theta&=&\frac{\lambda\ell}{\hbar\sqrt{{r^\star}^2+\frac{\lambda^2\ell^2}{\hbar^2}}},\nonumber\\
\sin\theta&=&\frac{r^\star}{\hbar\sqrt{{r^\star}^2+\frac{\lambda^2\ell^2}{\hbar^2}}}.
\end{eqnarray}
Differentiating the second of these with respect to time gives
\begin{equation}
\label{thetad}
\dot{\theta}=\frac{\lambda\ell\dot{r}^\star}{\hbar\sqrt{{r^\star}^2+\frac{\lambda^2\ell^2}{\hbar^2}}}.
\end{equation}
Combining this with 
\begin{equation}
\ell^2=\frac{m^2\lambda^2 r^2}{4f_2(r/\lambda)^2}\left(\dot{\theta}^2+\sin^2\theta\dot{\phi}^2\right),
\end{equation}
gives an equation for $\dot{\phi}$:
\begin{equation}
\label{phidot}
\dot{\phi}=\frac{2 f_2(r/\lambda)\ell}{r^\star m\lambda}\sqrt{1-\frac{m^2\lambda^4({\dot r}^\star)^2}{4 f_2(r/\lambda)^2\hbar^2\left({r^\star}^2+\frac{\lambda^2\ell^2}{\hbar^2}\right)}}.
\end{equation}
The final step is to eliminate ${\dot r}^\star$ from this equation.  For this we use (\ref{constb}) to write
\begin{equation}
\label{delta}
{\dot r}^2=\frac{4f_2(r/\lambda)^2r^2\hbar^2}{m^2\lambda^4}-\left(\frac{\lambda}{\hbar}\right)^2\left(e-\frac{\hbar^2r}{m\lambda^3}\left(f_1(r/\lambda)+f_3(r/\lambda)\right)-w(r)\right)^2-\frac{4f_2(r/\lambda)^2\ell^2}{m^2\lambda^2}\equiv\Delta(r).
\end{equation}
Using
\begin{equation}
\label{rad}
\dot r=\frac{{\dot r}^\star r^\star}{\sqrt{{r^\star}^2+\frac{\lambda^2\ell^2}{\hbar^2}}},
\end{equation}
and (\ref{rrstar}), we can write
\begin{equation}
{\dot r}^\star=\frac{\sqrt{\left({r^\star}^2+\frac{\lambda^2\ell^2}{\hbar^2}\right)\Delta(r^\star)}}{r^\star},
\end{equation}
with $\Delta(r^\star)\equiv\Delta\left(\sqrt{{r^\star}^2+\frac{\lambda^2\ell^2}{\hbar^2}}\right)$.  Substituting this in (\ref{phidot}) gives
\begin{equation}
\dot\phi=\frac{2 f_2(r^\star/\lambda)\ell}{r^\star m\lambda}\sqrt{1-\frac{m^2\lambda^4\Delta(r^\star)}{4 f_2(r^\star/\lambda)^2{r^\star}^2\hbar^2}}.
\end{equation}
Here $f_2(r^\star/\lambda)=f_2\left(\frac{1}{\lambda}\sqrt{{r^\star}^2+\frac{\lambda^2\ell^2}{\hbar^2}}\right)$.  From this the precession angle for half a cycle is easily obtained as
\begin{equation}
\Delta\phi=\int_{r_-^\star}^{r_+^\star}dr^\star\frac{2 f_2(r^\star/\lambda)\ell}{m\lambda\sqrt{\left({r^\star}^2+\frac{\lambda^2\ell^2}{\hbar^2}\right)\Delta(r^\star)}}\sqrt{1-\frac{m^2\lambda^4\Delta(r^\star)}{4 f_2(r^\star/\lambda)^2{r^\star}^2\hbar^2}}.
\end{equation}
Here $r_\pm^\star$ are the turning points. 

This form is still inconvenient as it is difficult to solve the energy and angular momentum in terms of the turning points $r_\pm^\star$.  It is much easier to solve them in terms of the turning points $r_\pm$. We therefore make a change of variables in the integral back to these quantities by using (\ref{rrstar}).  This yields
\begin{equation}
\label{prec}
\Delta\phi=\int_{r_-}^{r_+}dr\frac{2 f_2(r/\lambda)\ell}{m\lambda\sqrt{\left(r^2-\frac{\lambda^2\ell^2}{\hbar^2}\right)\Delta(r)}}\sqrt{1-\frac{m^2\lambda^4\Delta(r)}{4 f_2(r/\lambda)^2\hbar^2\left(r^2-\frac{\lambda^2\ell^2}{\hbar^2}\right)}}.
\end{equation}

The way we proceed is as follows: We choose turning points $r_+$ and $r_-$ and solve the energy, $e$, and angular momentum, $\ell$, from the conditions (see (\ref{delta}))
\begin{equation}
\Delta(r_+)=\Delta(r_-)=0.
\end{equation} 
There are two pairs of solutions, but the one pair has positive energy and complex angular momentum and is therefore unphysical.  The second pair has negative energy and real angular momentum and therefore describes elliptic motion.  The two solutions in this pair are simply related by a change  of sign of the angular momentum.  We do not list these expressions explicitly due to their length.

The resulting integral (\ref{prec}) cannot be performed exactly, but we can attempt an expansion of the integrant in orders of $\lambda$ and integrate term by term.  We do this for general functions $f_i(R)$, but it turns out to be convenient to write these functions in the following way:
\begin{equation}
f_i(R)=\frac{1}{R}h_i\left(\frac{1}{R}\right).
\end{equation}
The only restriction at this point on the functions $h_i(x)$ is that $h_i(0)=\frac{1}{2}$, which ensures the desired asymptotic behaviour as reflected in (\ref{zero}).  In the special case of (\ref{first}), it is easy to read off the explicit forms of these functions.  This long calculation yields for the precession over half a cycle
\begin{equation}
\label{precf}
\Delta\phi=\pm\left(\pi+\frac{ \pi  G M \left(r_-+r_+\right)\left(h_1^{\prime\prime}(0)-2h_2^{\prime\prime}(0)+h_3^{\prime\prime}(0)\right)}{4 r_- r_+\left(1+h_1^{\prime}(0)-2h_2^{\prime}(0)+h_3^{\prime}(0)\right)^2 }\left(\frac{ \lambda m}{ \hbar}\right)^2+O(\lambda,\hbar^0)+O(\lambda^2,\hbar^0)\right). 
\end{equation}
The sign depends on the choice of solutions, i.e. positive or negative angular momentum, which in turn depends on $1+h_1^{\prime}(0)-2h_2^{\prime}(0)+h_3^{\prime}(0)$ positive or negative.  For convenience we only consider the positive case from here on.  We note that if we take the limit $\lambda\rightarrow 0$ with $\hbar$ fixed, we recover the Newtonian result $\Delta\phi=\pi$.  Also note that we cannot take the $\hbar\rightarrow 0$ limit before the $\lambda\rightarrow 0$.  However, if we take the $\lambda\rightarrow 0$ and $\hbar\rightarrow 0$ limits such that $\frac{\lambda}{\hbar}$ is a fixed ratio, i.e 
\begin{equation}
\frac{\lambda}{\hbar}=\frac{1}{mv_0},
\end{equation}
where $v_0$ is the limiting speed of the non-commutative system, the terms $O(\lambda,\hbar^0)+O(\lambda^2,\hbar^0)$ vanish and we get
\begin{equation}
\label{precf1}
\Delta\phi=\pi+\frac{ \pi  G M \left(r_-+r_+\right)\left(h_1^{\prime\prime}(0)-2h_2^{\prime\prime}(0)+h_3^{\prime\prime}(0)\right)}{4 r_- r_+v_0^2\left(1+h_1^{\prime}(0)-2h_2^{\prime}(0)+h_3^{\prime}(0)\right)^2 }.
\end{equation}

Introducing the length of the semi-major axis $a=(r_+ +r_-)/2$ and the eccentricity $\epsilon=\frac{r_+-r_-}{r_++r_-}$, this reads
\begin{equation}
\label{precf2}
\Delta\phi=\pi+\frac{ \pi  G M \left(h_1^{\prime\prime}(0)-2h_2^{\prime\prime}(0)+h_3^{\prime\prime}(0)\right)}{2 a\left(1-\epsilon^2\right)v_0^2\left(1+h_1^{\prime}(0)-2h_2^{\prime}(0)+h_3^{\prime}(0)\right)^2 }.
\end{equation}

Substituting the form of the functions $h_i(R)$ as extracted from (\ref{first}), one obtains
\begin{equation}
\label{precf3}
\Delta\phi=\pi+\frac{ \pi  G M }{8 a\left(1-\epsilon^2\right)v_0^2 }.
\end{equation}
Note that since (\ref{precf2}) depends at most on the second order derivatives of $h_i(R)$, the inclusion of higher order terms for $f_2$ in (\ref{first}) cannot alter the result.  Remarkably, this result has the same form as the general relativistic (GR) result \cite{hartle}
\begin{equation}
\label{precGR}
\Delta\phi=\pi+\frac{3\pi GM}{c^2 a (1-\epsilon^2)},
\end{equation}
except for a numerical factor and the appearance of the limiting speed, rather then the speed of light. We discuss the physical ramifications of this result in section \ref{discussion}.

\subsection{Stable circular orbitals in a gravitational potential}

In this section we study the behaviour of stable circular orbits.  We make the following ansatz for these orbitals
\begin{equation}
x(t)=r\sin\theta\cos(\omega t),\quad y(t)=r\sin\theta\sin(\omega t), \quad z(t)=r\cos\theta.
\end{equation}
The only time dependence is therefore in the azimuthal angle $\phi$ that changes at a constant rate. 

This ansatz is inserted in the equations of motion (\ref{eqmf}) for the negative branch and with gravitational potential as in (\ref{gravpot}).  We first consider the equation of motion in the z-direction, from which one can solve for $\cot\theta$ in terms of the speed $V^2=\dot{\vec{X}}\cdot\dot{\vec{X}}$ as
\begin{equation}
\label{cot}
\cot \theta = \frac{V}{\sqrt{4 R^2 f_2(R)^2-V^2}}
\end{equation}

Using this result in the equation of motion for the x-and y-directions, which collapse to the same equation, one obtains the velocity as
\begin{equation}
\label{vel}
V=4 R f_2(R)\sqrt\frac{ab}{2(ab+c+d\sqrt{e})},
\end{equation}
where
\begin{eqnarray}
\label{vela}
a&=&\beta +R^2 \left(R \left(f_1'(R)-2 f_2'(R)+f_3'(R)\right)+f_1(R)-2 f_2(R)+f_3(R)\right),\nonumber\\
b&=&\beta +R^2 \left(R \left(f_1'(R)+2 f_2'(R)+f_3'(R)\right)+f_1(R)+2 f_2(R)+f_3(R)\right),\nonumber\\
c&=&4 R^4\left( f_2(R)^2- R^2 f_2'(R)^2\right),\nonumber\\
d&=&\beta +R^2 \left(R \left(f_1'(R)+f_3'(R)\right)+f_1(R)+f_3(R)\right),\nonumber\\
e&=&\beta ^2+R^2 \left(R \left(2 f_3'(R) \left(\beta +R^3 f_1'(R)\right)+R^3 f_1'(R)^2+2 \beta  f_1'(R)-16 R^2 f_2(R) f_2'(R)+R^3 f_3'(R)^2\right)\right.+\nonumber\\
&&\left.2 f_3(R) \left(\beta +R^3 \left(f_1'(R)+f_3'(R)\right)\right)+2 f_1(R) \left(\beta +R^2 \left(R \left(f_1'(R)+f_3'(R)\right)+f_3(R)\right)\right)+R^2 f_1(R)^2+R^2 f_3(R)^2\right).\nonumber\\
\end{eqnarray}
Substituting (\ref{vel}) in (\ref{cot}) gives $\cot\theta$ as a function of radius.  

These equations do not provide much insight into the behaviour of the velocity and $\cot\theta$ as a function of radius.  To simplify matters, we consider the long length scale behaviour in which we approximate the functions $f_i(R)$ as in (\ref{zero}).  After restoring dimensions using (\ref{dim}) and setting $\beta=\frac{GMm^2\lambda}{\hbar^2}-1\approx \frac{r_0}{\lambda}$ with $r_0=\frac{GM}{v_0^2}$ this gives 
\begin{eqnarray}
\label{vellong}
v(r)&=&v_0\sqrt{\frac{2}{1+\sqrt{1+4\left(\frac{r}{r_0}\right)^2}}},\nonumber\\
\cot\theta&=& \sqrt{\frac{2 }{\sqrt{1+4\left(\frac{r}{r_0}\right)^2}-1}}.
\end{eqnarray}

We note the following interesting behaviour
\begin{eqnarray}
\label{asymvel}
v(r)&=&v_0,\quad r<<r_0,\nonumber\\
v(r)&=&\sqrt{\frac{GM}{r}}, \quad r>>r_0.
\end{eqnarray}
Note, however, that the constant behaviour does not extend down to small radius as the short length scale corrections to the functions $f_i(R)$, which we neglected, become important and at short lengths scales one must consider the full expression (\ref{vel}). This result is intuitively simple to understand.  If there is a bounding speed, the dependence of the velocity on radius must be modified at small distances to avoid a violation of this limiting speed.  The only question is at what length scale this modification takes effect. We leave the discussion of the physical implications for section \ref{discussion}.

\section{Generalized dynamics}
\label{geneqm}

In this section we consider a generalisation of the results in the previous sections, based on the possible modification of the functions $f_i$ that appear in (\ref{b})-(\ref{d}).  
The source of such a modification of the functions $f_i$ relates to the choice of inner product on the quantum Hilbert space. In this regard it is crucial to realise that the choice of inner product and Laplacian is intimately connected by the requirement of hermiticity of the Laplacian.  Indeed, it can easily be checked that (\ref{eq:nc-laplacian}) is hermitian with respect to (\ref{innp}).  This choice of the inner product and Laplacian in turn determines the form of the functions $f_i(R)$ recorded in (\ref{b})-(\ref{d}).  If one changes the inner product from (\ref{innp}) to a more general form  
\begin{equation}
\label{innpm}
	(\psi|\phi)=4\pi\lambda^3{\rm tr}_c(\psi^\dagger f^2(\hat{R}/\lambda)\phi)\equiv 4\pi\lambda^3{\rm tr}_c(\psi^\dagger f^2(\hat{n}+1)\phi),
\end{equation}
for some non-vanishing, real function $f$,  the Laplacian (\ref{eq:nc-laplacian}) also needs to be changed to 
\begin{equation}
 \hat{\Delta}|\psi)=-|\frac{1}{\lambda^2f^2(\hat{R}/\lambda)}[\hat{a}^\dagger_\alpha,[\hat{a}_\alpha,\psi]])=|\frac{1}{\lambda^2f^2(\hat{n}+1)}[\hat{a}^\dagger_\alpha,[\hat{a}_\alpha,\psi]])
 \label{eq:nc-laplacianm}
\end{equation}
in order to maintain hermiticity.  Doing this one can quickly retrace the steps leading to the functions $f_i(R)$ given in (\ref{b})-({\ref{d}) to find that the modified functions are then
\begin{eqnarray}
f_1(R)&&=\frac{1}{2}\langle z_\alpha|\frac{1}{f^2(\hat{n}+2)}|z_\alpha\rangle,\label{bm}\\
f_2(R)&&=\frac{1}{2}\langle z_\alpha|\frac{1}{f(\hat{n}+1)f(\hat{n}+2)}|z_\alpha\rangle,\label{cm}\\
f_3(R)&&=\frac{1}{2}\langle z_\alpha|\frac{1}{f^2(\hat{n}+1)}|z_\alpha\rangle.\label{dm}
\end{eqnarray}
Note that (\ref{innp}) corresponds to the choice $f(x)=\sqrt{x}$.   This modification has two consequences: (1) The trace of the operator that projects on the subspace of spheres with radius $r\leq \lambda (\hat{n}+1)$ no longer yields the volume of a sphere in Euclidean space and (2) The dispersions relation of the free particle Schr\"odinger equation is modified.  Although this is acceptable at short length scales, these modifications are unwanted at long length scales and we therefore require that $f(x)$ has the asymptotic behaviour $f(x)\rightarrow\sqrt{x}$ when $x\rightarrow\infty$.  This generalisation may therefore be interpreted as introducing some form of curvature on configuration space, but such that it is asymptotically flat.  This provides a paradigm for a generalised interpretation of the equations of motion (\ref{eqmfa}) and constants of motion (\ref{const}) where the functions $f_i(R)$ are treated as generalised functions as in (\ref{bm})-(\ref{dm}).  Note that these function are not completely arbitrary, but that they are determined by one single function $f(x)$.

\section{Discussion}
\label{discussion}

We have now collected the most important results following from the non-commutative classical dynamics on fuzzy space. The challenge that remains is to extract a coherent physical picture from these results.  We discuss each result and its physical consequences separately.  Our discussion assumes a gravitational potential.  

\subsection{Limiting energy and speed}

One of the central features, which has cropped up on several occasions in the discussion above, is the existence of a cut-off energy with value $\frac{2 \hbar^2}{m\lambda^2}$, even for a free particle.  This result was also found in earlier studies of the quantum mechanics on fuzzy space where it essentially appears because the De Broglie wave-length cannot be made smaller then the non-commutative length scale.  The existence of an energy cut-off is certainly reasonable from the point of view of gravitational stability and one may hope that it may help to regulate ultra-violet divergencies in a fully fledged field theory.

Another striking result is the existence of a limiting speed $v_0=\frac{\hbar}{m\lambda}$.  The existence of such a limiting speed in itself is not so unconventional as we know that the speed of light also presents such a limit,  but rather its dependence on the mass of the object creates interpretational difficulties.  In the case of gravity the minus branch of the equation of motion (\ref{eqmfd}), which reduces to the Newtonian limit, reads
\begin{equation}
\label{eqmg}
\ddot{\vec{x}}=\frac{GM}{r^3}\left[\frac{m\lambda}{\hbar}\left(\vec{x}\times\dot{\vec{x}}\right)-\sqrt{1-\left(\frac{m\lambda}{\hbar}\right)^2\dot{\vec{x}}\cdot\dot{\vec{x}}}\;\vec{x}\right],
\end{equation}

If we assume that $\lambda$ is some fixed parameter only determined by the properties of space, this contradicts one of the most established principles in physics that motion under a gravitational force is independent of the mass, or indeed any other properties, of the test particle.  This is also the foundation for the geometrical interpretation of gravity as developed by Einstein and has been experimentally verified to great accuracy by the Dicke-E\"otv\"os experiment.  


It seems that the only way that (\ref{eqmg}) can be made consistent with observation, is if we assume that the limiting speed
\begin{equation}
\frac{\hbar}{m\lambda}=v_0,
\end{equation}
is a universal constant, independent from any properties of the test particle. This requires us to adopt the point of view that the commutation relations of the coordinates of a macroscopic particle with mass $m$ is given by 
\begin{equation}
\label{commc}
[\hat{x}_i,\hat{x}_j]=\frac{2\hbar}{mv_0}i\varepsilon_{ijk}\hat{x}_k.
\end{equation}
This implies that the properties of non-commutative space, or at least the coordinates of a massive particle moving in non-commutative space, must depend on the mass of the test particle, i.e. the non-commutative parameter must undergo some form of renormalization due to the presence of the test particle.  This is not a completely foreign notion as we know from GR that the local properties of space-time will be modified by the presence of a test particle.  It should also be noted that these commutation relations loose their validity on the microscopic level where $m$ is small.  In this case one must keep in mind that the equations of motion derived here under the saddle point approximation are invalidated as quantum fluctuations become important.  In this case a full quantum mechanical treatment is necessary as was done in \cite{scholtz5,press} where complete consistency with standard quantum mechanics was found.

It is at this point not clear what the value of $v_0$ should be.  It is, of course, tempting to adopt the speed of light as its value, but there are no compelling arguments for this as the equations of motion derived here only apply to massive particles and have nothing to say about the propagation of light.  The latter has to concern itself with the formulation of Maxwell's equations on fuzzy space.  Furthermore, as pointed out earlier, this limiting speed may have a general spatial dependence.  However, to avoid conflict with observation it must presumably be a substantial fraction of the speed of light at terrestrial or solar scales.  Indeed, the precession obtained in (\ref{precf3}) suggests that  $v_0=\frac{c}{\sqrt{24}}$, (approximately 20\% of the speed of light) to obtain agreement with observation for Mercury.  It should, however, be noted that the flat space approximation used in its derivation may be inadequate and that the inclusion of curvature, as described in section \ref{geneqm}, may resolve the numerical conflict between (\ref{precf3}) and the GR result (\ref{precGR}), resulting in $v_0=c$. On galactic scales the situation is seemingly different.  Agreement between the predicted constant velocity curves of (\ref{vellong}) and its asymptotic behaviour (\ref{asymvel}) with the observed velocity curves, requires values of $v_0$ much less than the speed of light.  We turn to this in the next subsection.  

At this point the appearance of this limiting speed remains problematic and understanding its role within a consistent physical paradigm remains a challenge.  One obvious way of avoiding the issue and any possible observational conflict is to adopt the speed of light as its value, but even the implications of this require a careful analysis.  If this is done, the resulting dynamics deviates only slightly from standard Newton dynamics.  In particular it sheds no new light on the behaviour of velocity curves as the limiting speed is too large and the length scales on which the curves are constant too short.  Although disappointing, it may turn out to be the only consistent physical paradigm.

\subsection{Velocity curves}

One of the most attractive features of the current dynamics is the flatness of the velocity curve below the length scale $r_0=\frac{GM}{v_0^2}$.  However, as pointed out above, this can only have observational relevance if the limiting velocity is much less then the speed of light and most likely also spatially dependent.  Both these features are unconventional.   Nevertheless, let us run through this scenario as the enigma of observed velocity curves warrants unconventional approaches.  

Let us therefore assume the existence of a limiting speed much less than the speed of light (of the order of 100-300 km.s$^{-1}$) in the non-commutative scenario described above.  If one moves out from the center of a galaxy, the velocity curve grows simply because the included mass grows and several models for this exist \cite{soufe}.  However, once the velocity reaches the limiting velocity, the curve must saturate at this value and stay there up to the length scale $r_0$, after which it will assume the standard $\sqrt{\frac{GM}{r}}$ behaviour as in (\ref{asymvel}).  This does, of course, require us to take the limiting velocity as the plateau velocity, commonly denoted $v_f$, of the observed velocity curve.   The important point to realise though is that in this scenario flat velocity curves are natural and indeed generic and no specific distribution of the mass in the galaxy has to be assumed for flatness.  In fact, one may assume that all the mass is concentrated at the center as we indeed did when deriving (\ref{vellong}).

Figure \ref{velcurve} shows data for the Milky way from \cite{batt} with galactic constants $R_0=8$ kpc and $V_0=200$ km.s$^{-1}$ up to 200 kpc.  We also show a least squares fit of the velocity curve (\ref{vellong}) to the data (solid line).  This gives $r_0=76.5$ kpc and $v_0=215.6$ km.s$^{-1}$, which requires a galactic mass of $8.3\times 10^{11} M_\odot$.  This is in complete agreement with the mass $M(200{\rm kpc})=6.8\pm 4.1\times 10^{11} M_\odot$ reported in \cite{batt}, but larger then the mass $M(100{\rm kpc})=3\times 10^{11} M_\odot$ reported in \cite{soufe}.  Note, however, the difference in radius so that a larger value is to be expected.  Let us also consider what happens if we just consider the baryonic mass (stellar and gaseous). A reliable estimate of this can be obtained from the empirical baryonic Tully-Fisher relation \cite{smac}
\begin{equation}
a_0 G M=v_f^4.
\end{equation}
We take for the constant $a_0$ the value reported in \cite{smac} of $a_0=1.3\times 10^{-10}$ m.s$^{-2}$ and $v_f=200$ km.s$^{-1}$. Using these values we find the baryonic mass of the Milky way to be $9.3\times 10^{10} M_\odot$.  This gives $r_0=9.96$ kpc for $v_0=v_f=200$ km.s$^{-1}$.  This velocity curve is also shown as the dotted line in figure \ref{velcurve}.  Clearly, the value of $r_0$ is too small if only baryonic mass is considered to explain the extent of the plateau observed in the velocity curve.  Indeed, as usual, we see that the baryonic mass only makes up around 11\% of the galactic mass required to explain the data.  However, as mentioned before, in this scenario no assumptions about the distribution of this excess mass in the galaxy needs to be made to explain the flatness of the velocity curve.  It may therefore even be possible that this mass is concentrated in the center of the galaxy, e.g. in the form of a massive black hole.  One may be concerned that this concentration of mass may be detectable through the motion of nearby stars such as S2 \cite{S2} and this would certainly be the case in a Newtonian paradigm.  However, in the current paradigm the limiting speed may prevent such a detection if it is low enough and application of a Newtonian paradigm will lead to an underestimation of the mass.  In fact, in the current paradigm the only way the included mass can be estimated accurately is through the length scale $r_0$ as the velocity is largely independent from the included mass below this scale.  This does, however, pose a further difficulty.  The speed of S2 at its perihelion is around 5000 km.s$^{-1}$ \cite{S2}, much larger than the plateau value of the velocity curve of around 200 km.s$^{-1}$.  If all the excess mass is concentrated at the center of the galaxy, it requires us to assume that the limiting velocity must be spatially dependent, which can be accommodated through a generalized function $f_2(r)$ as described in section \ref{geneqm}.

Finally, we must establish that the velocity curves predicted by this paradigm does not conflict with observation on the solar scale where the Newtonian paradigm clearly holds.  To do this we take for $v_0$ the value consistent with the precession of Mercury, i.e. $v_0=\frac{c}{\sqrt{24}}$.  Computing the length scale $r_0$ for the mass of the sun and this limiting speed gives $r_0=35 417$ km, which still falls inside the sun.  From this we expect the standard $v=\sqrt{\frac{GM}{r}}$ behaviour for the rotation velocities on the solar scale and no observational conflict.

 \begin{figure}[t]
   \centering
   \includegraphics[width=0.8\textwidth]{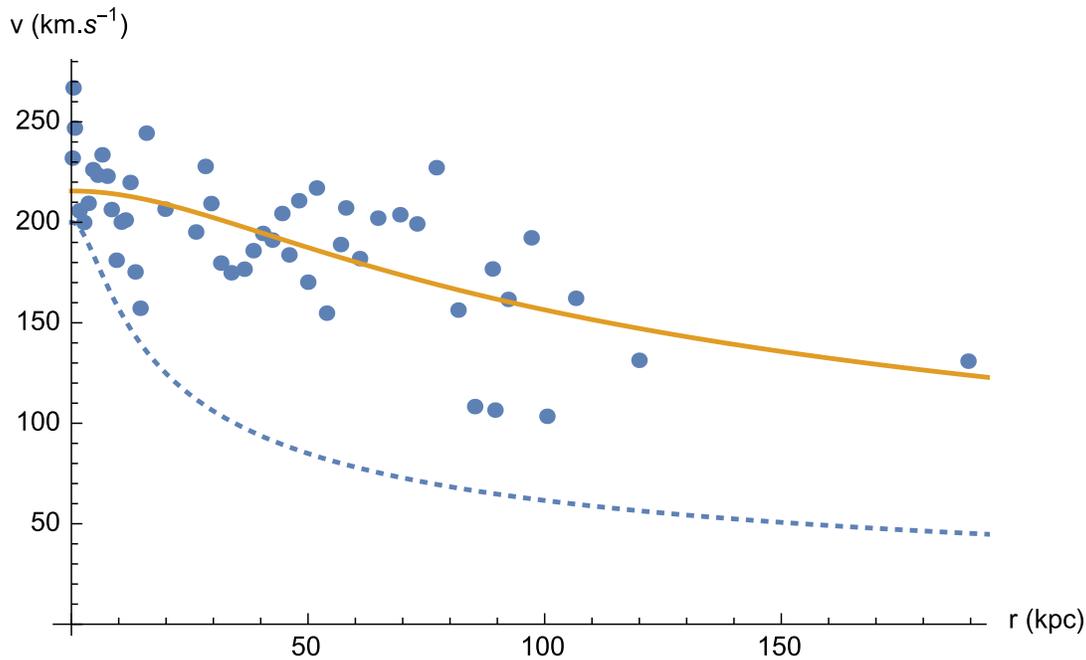}
     \caption{Velocity curve and data for the Milky way.  The data is from \cite{batt} with galactic constants $R_0=8$ kpc and $V_0=200$ km.s$^{-1}$.  The dashed curve is for an estimated baryonic mass of $9.3\times 10^{10} M_\odot$ and $v_0=200$ km.s$^{-1}$, which gives $r_0=9.96$ kpc. The solid line is a least squares fit of the velocity curve (\ref{vellong}) to the data with mass $8.3\times 10^{11} M_\odot$ and $v_f=v_0=215.6$ km.s$^{-1}$}
  \label{velcurve}
 \end{figure}

\subsection{Two branches}

One of the features of the equations of motion derived above is the existence of two branches with disconnected dynamics of which only one reduces to standard Newtonian dynamics in the commutative limit.  The choice of branch is determined by the value of the energy as reflected in (\ref{constmd}) and, since energy is conserved, is set by the initial conditions.  In particular, we note that for $\frac{\hbar^2}{m\lambda^2}<e<\frac{2 \hbar^2}{m\lambda^2}$ or, equivalently, $\frac{\hbar  v_0}{\lambda}<e<\frac{2 \hbar v_0}{\lambda}$, the dynamics must be described by the plus branch, which does not reduce to Newtonian dynamics in the commutative limit.  Note, however, from (\ref{constmd}) that high velocities are not required for this energy range.  On the unified level of the effective potential for the radial motion discussed in section \ref{genorb}, the presence of these two branches manifests itself in that the effective potential becomes completely repulsive for this range of energies (\ref{effplong}).   This implies that a test particle with energy in this range will indefinitely accelerate radially outwards.

\subsection{Experimental testability}

A feature that sets the current dynamics apart from Newtonian or general relativistic dynamics is the offset of the source of the gravitational potential from the plane of motion.  This is encoded in (\ref{press1}) where the offset of the angle $\theta$ from $\frac{\pi}{2}$ measures this displacement.  One can, of course, also express it in terms of the absolute distance $r\cos\theta$ appearing in (\ref{press1}).  Assuming the asymptotic form (\ref{zero}) for the functions $f_i$, one can easily check that the magnitude of the angular momentum in (\ref{press1}) coincides with the commutative result, i.e. $\vec{\ell}=m\left(\vec{x}\times\dot{\vec{x}}\right)$.  Note, though, that this is only true for the magnitude and not the individual components of the commutative angular momentum, which are in fact not even conserved.  We have a simple result for the magnitude of the commutative angular momentum in terms of the parameters of the elliptic orbit of the test particle:
\begin{equation}
\ell=m\sqrt{\frac{2GMr_+r_-}{r_++r_-}}.
\end{equation}
Using this in (\ref{press1}), we obtain for the displacement $d$, which is also a constant of motion
\begin{equation}
d=r\cos\theta=\frac{1}{v_0}\sqrt{\frac{2GMr_+r_-}{r_++r_-}}.
\end{equation}
Hence, this displacement gives a direct measure of $v_0$.  

If we take $v_0$ to have the value suggested by the precession of Mercury, i.e. $v_0=\frac{c}{\sqrt{24}}$ gives $d=44 301$ km, which is a small fraction of the sun's radius.  In terms of the angle $\theta$, this implies a deviation of $9.63\times 10^{-4}$ rad at the perihelion and $6.34\times 10^{-4}$ rad at the aphelion.  Measurement of this offset can falsify the current non-commutative dynamics and, if a non-zero value is found, it can give a direct measure of the limiting velocity $v_0$.  

\section{Summary and conclusions}
\label{summary}

We have derived the path integral action for a particle moving in fuzzy space and the corresponding classical equations of motion.  The main features of these equations are a cut-off  energy, a generally spatial dependent limiting speed, planetary precession remarkably similar to the general relativistic result, velocity curves that plateau below the length scale $\frac{GM}{v_0^2}$, displaced planar motion and the existence of two dynamical branches of which only one reduces to Newtonian dynamics in the commutative limit.  The branch that does not reduce to Newtonian dynamics predicts a repulsive effective potential for the radial motion and indefinite outward acceleration.  Most of these features are unconventional and observational data pose a severe challenge for this scenario.  Yet, they also offer attractive features that may have relevance for the enigmas of dark matter and dark energy.  

Avoiding conflict with the Dicke-E\"otv\"os experiment requires the adoption of mass dependent commutation relations for coordinates and a universal limiting speed.  The role of this limiting speed is not yet understood and it may play a key role in developing a consistent physical paradigm for the results reported above.  In particular, a rather conventional paradigm results if the limiting speed is set equal to the speed of light and only a slight modification of Newton dynamics can be expected.  These modification may, however, still be detectable through sensitive experiments.

\end{document}